\documentclass[12pt,showkeys,prab,superscriptaddress,floatfix,nofootinbib,longbibliography]{revtex4-2}

\usepackage{graphicx}
\usepackage{color}
\usepackage{amsfonts}
\usepackage{bm}
\usepackage{xspace}
\usepackage{amsmath}

\newcommand {\bfv} {{\bf v}}

\renewcommand {\d} {{\rm d}}

\newcommand {\E} {\varepsilon}

\newcommand {\om} {\omega}
\newcommand {\Om} {\Omega}

\newcommand {\lamu} {\lambda_{\rm u}}

\newcommand{\MBNExplorer} {\textsc{MBN Explorer}\xspace}
\newcommand{\MBNStudio}{\textsc{MBN Studio}\xspace}

\newcommand{\calA} {{\cal A}}

\begin{document}

\title{\textcolor{black}{Electron and positron channeling and photon emission processes in boron doped periodically bent diamond}}


\author{Andrei V. Korol}
\email[]{korol@mbnexplorer.com}
\affiliation{MBN Research Center, Altenh\"{o}ferallee 3, 60438 Frankfurt am Main, Germany}

\author{Andrey V. Solov'yov}
\email[]{solovyov@mbnresearch.com}
\affiliation{MBN Research Center, Altenh\"{o}ferallee 3, 60438 Frankfurt am Main, Germany}

\begin{abstract}

In this paper, theoretical and numerical analyses
are conducted of the profiles of the planar (-110) crystallographic
direction in the diamond layer doped with boron atoms.
The planar profiles for periodic doping following several
ideal dependencies of the boron concentration on the distance in the crystalline medium.
%
Numerical simulations of the channeling and photon emission processes
have been carried out for 855 MeV electron and 530 MeV positron beams
incident on boron-doped diamond with a four-period bending profile
in the samples grown at the European Synchrotron Radiation Facility
(ESRF).
The simulations were performed using the MBN Explorer software package.
It is shown that the channeling efficiency and the intensity of the
crystalline undulator radiation strongly depend on the orientation of
the incident beam relative to the bent channel profile at the
entrance to the boron-doped layer.
For the same conditions at the crystal entrance,
the intestity of
radiation emitted by positrons is significantly higher than that for
electrons.


\end{abstract}

\maketitle

\section{Introduction \label{Introduction}}

The interaction of high-energy charged particles with
crystals is sensitive to the incoming beam direction relative
to the target's major crystallographic directions.
Projectiles, impacting upon a crystal at small angles to its planes
(or axes) can propagate large distances in the crystalline media
following the planar (axial) direction.
This specific motion, termed channeling motion, is due to the
collective action of the electrostatic fields of the lattice atoms
\cite{Lindhard}.
Research into the channeling process of ultra-relativistic projectiles
in oriented crystals has emerged as a broad field
\cite{BiryukovChesnokovKotovBook,Uggerhoj:RPM_v77_p1131_2005,%
ChannelingBook2014,CLS-book_2022} with applications encompassing
beam steering \cite{MazzolariEtAl:EPJC_v78_p720_2018},
collimation \cite{AfoninEtAl:PRL_87_094802_2001},
focusing \cite{ScandaleEtAl:PRAB_v21_014702_2018}
and extraction \cite{SytovEtAl:EPJC_v82_187_2022}.

Oriented crystals of different geometries irradiated by
beams of ultra-relativistic electrons and positrons can potentially
serve as novel crystal-based light sources (CLS) that produce intense
gamma-ray radiation in the MeV-GeV photon energy range
\cite{CLS-book_2022,SushkoKorolSolovyov:EPJD_v76_166_2022,%
KorolSolovyov:NIMB_v537_p1_2023}.
As demonstrated recently
\cite{KorolSushkoSolovyov:PRAB_v27_p100703_2024},
the photon fluxes generated in such systems can exceed the levels
achievable at modern laser-Compton gamma-ray light sources.
The practical realization of CLSs is the subject of the ongoing
Horizon Europe EIC-Pathfinder-2021 project TECHNO-CLS
\cite{TECHNO-CLS}.

One of the proposed schemes for the implementation of CLSs
employs crystalline undulators (CU), wherein the radiation is
emitted by ultra-relativistic electrons or positrons
channeling in a periodically bent crystal
\cite{KSG1998,KSG_review2004,
Tabrizi-EtAl:PRL_v98_164801_2007,ChannelingBook2014}.
In such systems, in addition to the channeling radiation
\cite{ChRad:Kumakhov1976}, the CU radiation (CUR) is emitted
due to the periodicity of the trajectory of a particle following the
bending profile \cite{KSG1998,KSG_review2004}.
PBCs are advantageous because the  characteristics of the emitted
radiation can be optimised for given parameters of the incident beam
by varying the amplitude $a$ and the period $\lamu$ of the bending.
However, the operating efficiency of such a device depends
on the quality of periodic bending \cite{ChannelingBook2014}.
The fabrication of PBCs, which are typically made of materials such as
silicon, germanium, or diamond, poses a significant technological
challenge.
In order to achieve the periodical deformation of the crystalline
structure, several techniques have been developed.
These include mechanical scratching \cite{BellucciEtAl:PRL_v90_034801_2003},
laser ablation technique \cite{Balling_EtAl-NIMB_v267_p2952_2009},
the grooving method
\cite{Guidi_etal:NIMB_v234_p40_2005,Bagli_etal:EPJC_v74_p3114_2014},
tensile/compressive strips deposition
\cite{Guidi_etal:APL_v90_114107_2007,
Guidi-EtAl_ThinSolFilms_v520_p1074_2011,
MalaguttiEtAl:NIMA_v1076_170480_2025},
sandblasting \cite{Camattari-EtAl:JApplCryst_v50_p145_2017},
ion implantation \cite{Bellucci_EtAl-APL_v107_064102_2015},
and ultrasonic waves
\cite{KSG1998,KSG_review_1999,BaryshevskyDubovskayaGrubich1980,%
KalerisEtAl:PRAB_v28_033502_2025}.

Another approach to achieving periodic bending in crystals was
proposed \cite{Breese:NIMB_v132_p540_1997,
MikkelsenUggerhoj:NIMB_v160_p435_2000,%
AvakianEtAl-NIMA_v508_p496_2003}, which involves the introduction of
dopant atoms to the crystalline structure.
This method enables the reduction of the bending period to a range
of a few microns.
The fabrication of doped crystals can be achieved through a
variety of methods, with each technique suited to growing
different types of crystals.
In the case of CLSs, particular attention has been focused on two
crystal types: Silicon crystals doped with Germanium
(Si$_{1-x}$ Ge$_{x}$), grown through molecular beam epitaxy
\cite{Sakai_2013},
and Diamond crystals doped with Boron (C$_{1-x}$ B${_x}$), grown via chemical vapour deposition (CVD) \cite{ConnellEtAl_CVD_2015}.
Here, $x$ denotes the dopant relative concentration.
The boron-doped diamond layer cannot be separated from a
single diamond substrate on which the superlattice is synthesised.
Therefore, unlike Si$_{1-x}$ Ge$_{x}$ superlattice,
a diamond-based superlattice essentially has a heterocrystal structure,
i.e., it consists of two segments: a straight single-crystal diamond
substrate and a periodically doped layer.

In recent years, several channeling experiments have been carried out
with periodically bent doped crystals to detect CUR
\cite{Backe_EtAl:NuovCimC_v34_p157_2011,Backe_EtAl_2013,%
BackeEtAl:JPConfSer_v438_012017_2013,Wistisen_EtAl:PRL_v112_254801_2014,
UggerhojWistisen:NIMB_v355_p35_2015,Wistisen_EtAl:EPJD_v71_124_2017,
BackeEtAl:arXiv_2404.15376v3}
but no clear CUR signal was detected.
The first observation of the CUR peak has been reported very recently
\cite{BackeEtAl:arXiv_2504.17988} for a 855 MeV electron beam
channeling in a periodically bent oriented diamond(110) crystal.
We note that the existance of this peak, its position and intensity
were predicted earlier in Refs.
\cite{PavlovEtAl:EPJD_v74_21_2020,PavlovEtAl:SPB_v14_190_2021}.

In this paper we carry out theoretical analysis of the profile of
periodically bent planes in the diamond layer doped periodically with
boron atoms.
It is demonstrated that it is essential to know the
precise orientation of the undulator axis relative to the
crystallographic directions in the substrate.
This knowledge is needed to determine
(i) the optimal direction of the incident beam,
and
(ii) the position of a detector to register the emitted radiation.
The most intense undulator radiation is emitted in the
cone $\theta_0\sim 1/\gamma$ along the undulator axis
(where $\gamma$ is the Lorentz factor of the ultra-relativistic
charged particle).
Therefore, to maximise the collection of the radiation the detector
must be placed on the axis.
This is particularly important if the detector aperture is smaller than
the natural emission cone $1/\gamma$.
The results of numerical simulations of
the radiation emitted by
ultra-relativistic electrons and positrons presented in Section
\ref{CaseStudies} support this statement.

The simulations of the particles passage in a crystalline media and
of the photon emission process have been performed using the \MBNExplorer
software package \cite{MBNExplorer_2012} for advanced multi-scale
modelling of complex molecular structures and dynamics.
\MBNStudio \cite{SushkoEtAl_2019-MBNStudio},
a multi-task toolkit and a dedicated graphical
user interface for \MBNExplorer, was used to construct the systems,
prepare input files, and analyse simulation outputs.
A special module of the package \cite{MBN_ChannelingPaper_2013}
allows the simulation of the passage of various particles (positively
and negatively charged, light and heavy) through a variety of media,
including single, bent and periodically bent crystals,
hetero-crystalline structures (such as superlattices), and many others.
The software simulates the trajectories within the framework of
classical Relativistic Molecular Dynamics (RelMD).
The dedicated computational algorithms allow the simulations of
particle dynamics at macroscopically large distances and radiation emission from propagating projectiles with atomistic accuracy.
The module has been extensively used to simulate the propagation of
ultra-relativistic charged particles in oriented crystals, accompanied
by the emission of intense radiation.
A comprehensive description of the RelMD approach implemented in
\MBNExplorer as well as a number of case studies are presented
in a review article \cite{KorolSushkoSolovyov:EPJD_v75_p107_2021}.

The general methodology implemented in MBN Explorer has been used
to generate trajectories of ultra-relativistic electrons and positrons
in a crystalline environment the randomness accounts for randomness
in the sampling of the incoming projectiles as well as in the
displacement of the lattice atoms from their nodal positions due to
thermal vibrations.
As a result, each trajectory corresponds to a unique crystalline
environment.
Another phenomenon that contributes to the statistical independence of
the simulated trajectories has been taken into account.
This concerns the events of inelastic scattering of a projectile
particle from the crystal atoms resulting in the atomic excitation or
ionisation.
These collisions lead to a random change in the velocity of
a projectile.
These quantum events are random and occur on the atomic scale
in terms of time and space, therefore, they are be incorporated
into the classical mechanics
framework in according to their probabilities \cite{SushkoEtAl:arXiv_2404.15376v3}.

In Section \ref{Initial} the formalism that allows one to
quantify the change in the lattice constant due to the addition of
the dopant atoms \cite{Breese:NIMB_v132_p540_1997,PicrauxEtAl:NIMB_v33_p891_1988,
DeSalvadorEtAl:PhysRevB_v61_p13005_2000,KKSG:NIMA_v483_p455_2002,
Backe_EtAl_2013}
is applied to analyse the geometry of the
profiles of the bent planes in the boron doped layer for several
schemes of periodic and quasi-periodic doping.
The results of the numerical simulations are presented and
discussed in Section \ref{CaseStudies}.
The exemplary case study considered is related
to 855 MeV electrons and 530 MeV positrons channeling
along the (-110) planar direction in a four-period boron doped periodically bent diamond fabricated at ESRF and used in recent
experiments with the electrons at the MAinz MIkrotron (MAMI) facility
\cite{BackeEtAl:arXiv_2404.15376v3}.
Section \ref{Conclusion} summarizes the results of this work and
presents future perspectives.

\section{Theoretical framework  \label{Initial}}

In this Section theoretical analysis is presented that
that relates the parameters of periodic bending of crystal planes
to the dopant concentration that varies along
a particular crystallographic direction.

\begin{figure} [h]
\centering
\includegraphics[clip,width=15cm]{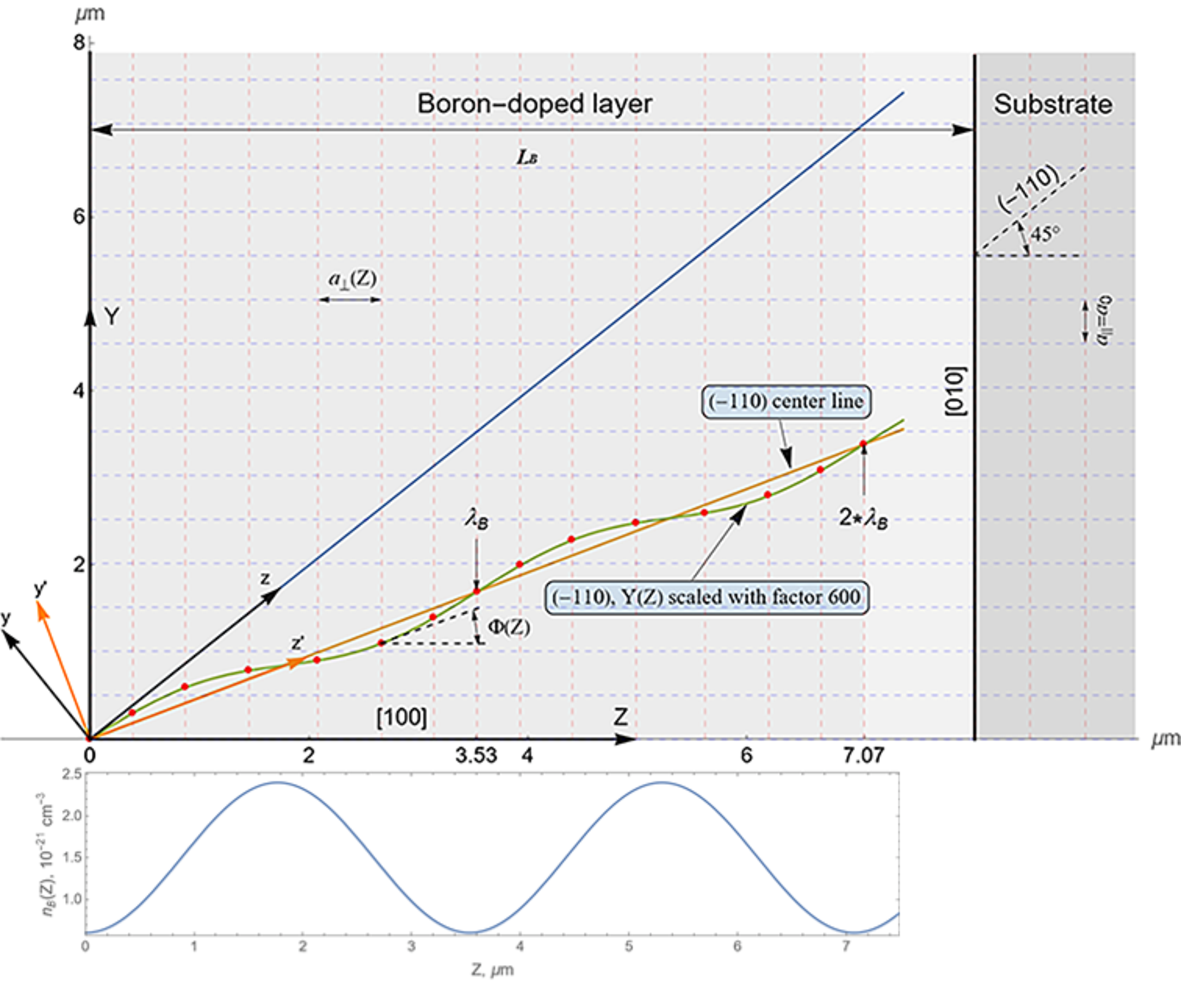}
\caption{
Sketch of the deformation of the crystal structure
in a boron doped diamond layer of thickness $L_{\rm B}$
grown on a (100) diamond substrate.
In the substrate, the lattice constant is the same along the
directions parallel and perpendicular to the (100) plane:
$a_{\|}=a_{\perp}=a$.
The addition of the dopant atoms increases the lattice constant
$a_{\perp}(Z)$ along the [100] axial direction (note that
the $Z$ axis is directed towards the substrate).
As a result, the angle $\Phi$ between the (-110) plane and
the [100] axis, which is equal to 45$^{\circ}$ in the substrate,
becomes $Z$ dependent so that the plane profile $Y(Z)$ deviates
from a straight line shown as the $z$ axis.
Two periods (along the $z$ axis) of the profile shown
correspond to cosine-like periodic dependence of the boron
concentration $n_{\rm B}(Z)$ with the period
$\lambda_{\rm B}=3.53$ $\mu$m (drawn at the
bottom of the figure).
}
\label{Figure01.fig}
\end{figure}

As a case study we consider a heterocrystal consisting of two parts:
(1) the oriented single diamond (100) substrate and (2) the diamond
crystal doped with boron atoms whose concentration $n_{\rm B}$ varies
along the crystallographic axis $[100]$ (the $Z$ direction), see Fig.
\ref{Figure01.fig}.
This causes the lattice constant $a_{\perp}$ to vary along $Z$.
In the $Y$ direction (the $[010]$ crystallographic axis) the in-plane
lattice constant $a_{\|}$, i.e. the one measured along the $[010]$ and
$[001]$ crystallographic directions, does not change, being equal to
its equilibrium value $a_0=3.567$ \AA{} in the diamond crystal.
Following Refs.
\cite{ConnellEtAl_CVD_2015,WojewodaEtAl-DiamondRelMat_v17_p1302_2008,
BrazhkinEtAl-PRB_v74_140502_2006,BrunetEtAl-DiamondRelMat_v7_p869_1998}
one considers the linear dependence of  $a_{\perp}$
on the boron concentration:
\begin{eqnarray}
a_{\perp}(Z)
=
a_0\left(1 + \kappa n_{\rm B}(Z) \right)\,.
\label{Methodology:eq.05}
\end{eqnarray}
The values of the coefficient $\kappa$, calculated using the
Vegard's law \cite{Vegard:ZPhysik_v5_p17_1921} and in the linear approximation for average atomic volume \cite{BrazhkinEtAl-PRB_v74_140502_2006} are as follows:
\begin{eqnarray}
\kappa\, \left[\mbox{cm$^{3}$}\right]
=
10^{-25} \times
\left\{
\begin{array}{ll}
8.12, &\mbox{Ref. \cite{Vegard:ZPhysik_v5_p17_1921}}\\
5.38, &\mbox{Ref. \cite{BrazhkinEtAl-PRB_v74_140502_2006}}\\
\end{array}
\right.
\label{K_coefficient:eq.02}
\end{eqnarray}
For brevity, the $\kappa$ values calculated according to
Refs. \cite{BrazhkinEtAl-PRB_v74_140502_2006} and \cite{Vegard:ZPhysik_v5_p17_1921} are notated below as
$\kappa_{\mbox{\footnotesize\cite{BrazhkinEtAl-PRB_v74_140502_2006}}}$
and
$\kappa_{\mbox{\footnotesize\cite{Vegard:ZPhysik_v5_p17_1921}}}$,
respectively.
Other quantities that depend on $\kappa$ are similarly notated.

Note that $\kappa_{\mbox{\footnotesize\cite{Vegard:ZPhysik_v5_p17_1921}}}$
and $\kappa_{\mbox{\footnotesize\cite{BrazhkinEtAl-PRB_v74_140502_2006}}}$
were used in Refs. \cite{ConnellEtAl_CVD_2015} and \cite{BackeEtAl:arXiv_2404.15376v3},
respectively.

In boron-doped diamonds grown by the MPCVD method
the concentration of the boron atoms does not exceed a few percent
of the concentration $n=1.763\times10^{23}$ cm$^{-3}$
of carbon atoms in the diamond crystal
\cite{ConnellEtAl_CVD_2015,ThuNhiTranThi_JApplCryst_v50_p561_2017}.
Therefore, a strong inequality
\begin{eqnarray}
\kappa n_{\rm B}(Z)\ll 1
\label{Methodology:eq.07}
\end{eqnarray}
is implied below.

The change in $a_{\perp}$ leads to a distortion of the
crystallographic directions that are not parallel to
the $(100)$ plane.
In particular, the angle $\Phi$ between the $(-110)$ plane
(shown by the blue line in the figure) and the $Z$ axis deviates from
its value of $45^{\circ}$ in the single crystal (the dashed line).
In the doped crystal the angle changes with $Z$ so that
\begin{eqnarray}
{\d Y \over \d Z}
=
\tan \Phi(Z)
=
{a_0 \over a_{\perp}(Z)}
\approx
1 - \kappa n_{\rm B}(Z)\,.
\label{Methodology:eq.03}
\end{eqnarray}
Hence, the profile $Y=Y(Z)$ of the deformed $(-110)$ plane
is calculated as follows:
\begin{eqnarray}
& &
Y(Z) =
Z - \Delta_Z(Z),
\label{Methodology:eq.09a} \\
\mbox{with}
&&
\Delta_Z(Z)
= \kappa \int_0^Z n_{\rm B}(Z^{\prime})\,\d Z^{\prime}\,.
\label{Methodology:eq.09}
\end{eqnarray}
Because of the condition (\ref{Methodology:eq.07}),
the strong inequality $\Delta_Z(Z)/Z \ll 1$ is
valid for all values of $Z$.
However, it is important to know the explicit dependence
$\Delta_Z(Z)$ in order to choose (i) the direction of the incident beam
that provides the optimal conditions for channeling, and
(ii) the orientation of the cone within which the intensity
of the emitted radiation is maximised.

For further reference let us write the profile
(\ref{Methodology:eq.09a}) in terms of the coordinates $(y,z)$
where
the $y$ axis is aligned with the [-110] axial
direction and the $z$ axis lies within the (-110) plane in the substrate, see Fig. \ref{Figure01.fig}.
Using the approximation
$z=\sqrt{2}Z- \Delta_Z(Z)/\sqrt{2} \approx \sqrt{2}Z$ one obtains
\begin{eqnarray}
 y(z)
 \approx
-
{\Delta_Z(z/\sqrt{2}) \over \sqrt{2}}\,.
\label{Methodology:eq.12}
\end{eqnarray}
Note that thickness $L_{\rm u}$ of the doped layer along the $z$
direction is calculated as $L_{\rm u}\approx \sqrt{2}L_{\rm B}$ where
$L_{\rm B}$ is the thickness along the $Z$ axis,
see Fig. \ref{Figure01.fig}.

Eqs. (\ref{Methodology:eq.09}) and (\ref{Methodology:eq.12}) allow
the bending profile to be derived for any  dependence
$n_{\rm B}(Z)$ of the dopant concentration.
The data on $n_{\rm B}(Z)$ can be collected in the process of
the doped-diamond growing.

Firstly, derive explicit expressions for the profiles corresponding
to several ideally periodic boron doping dependencies $n_{\rm B}(Z)$,
in which the dopant concentration varies from
$n_{\rm B}^{\min}$ to $n_{\rm B}^{\max}$ with
fixed period $\lambda_{\rm B}$ measured along the $Z$ axis:
\begin{subequations}
\label{allequations}
\begin{eqnarray}
n_{\rm B}(Z)
 =
 n_{\rm B}^{\min}
 +
 {n_0 \over 2}\left(1 - \cos{2\pi Z \over \lambda_{\rm B}}\right)
 &\mbox{\quad - "cosine" doping},
\label{CaseStudies:eq.01a} \\
n_{\rm B}(Z)
 =
 n_{\rm B}^{\min}
 +
 {n_0 \over 2}\left(1 \textcolor{blue}{-} \sin{2\pi Z \over \lambda_{\rm B}}\right)
 &\mbox{\quad - "sine" doping},
\label{CaseStudies:eq.01b}\\
n_{\rm B}(Z)
=
n_{\rm B}^{\min}
+
2n_0
\left\{
\begin{array}{ll}
\displaystyle
{Z \over\lambda_{\rm B}} -k,
& \mbox{for}\ {Z\over\lambda_{\rm B}}=[k, k+0.5]  \\
\displaystyle
k+1 - {Z \over\lambda_{\rm B}},
& \mbox{for}\ {Z\over\lambda_{\rm B}}=[k+0.5, k+1]
\end{array}
\right.
 &\mbox{\quad - "saw" doping}.
\label{CaseStudies:eq.01c}
\end{eqnarray}
\end{subequations}
Here $n_0 = n_{\rm B}^{\max} - n_{\rm B}^{\min}$ and
in the last equation $k=0,1,2,\dots$.

\begin{figure} [h]
\centering
\includegraphics[clip,width=13cm]{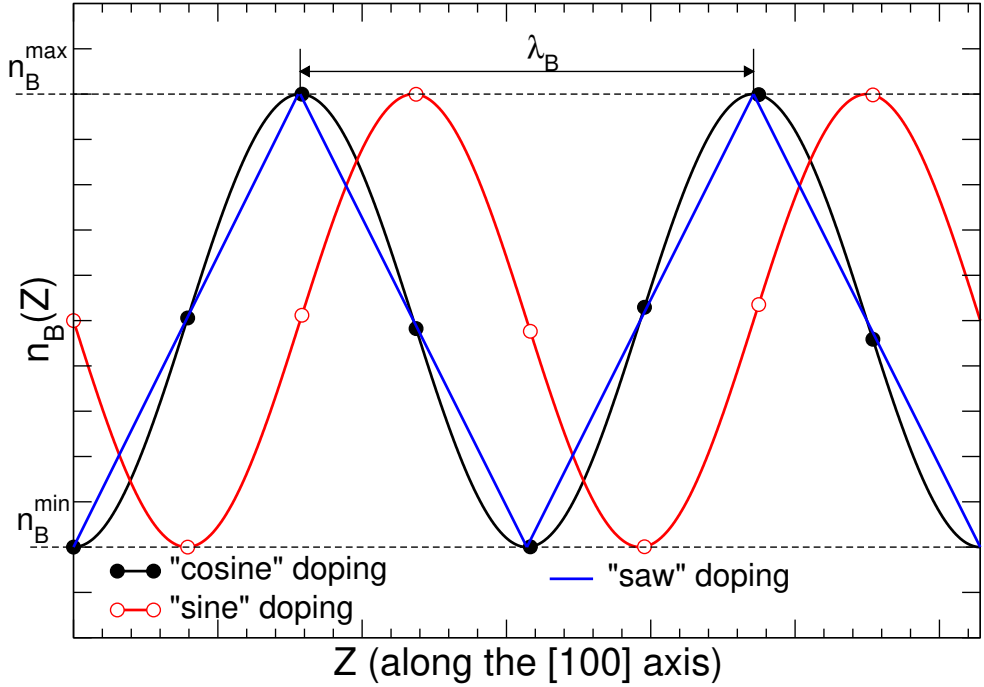}
\caption{
First two periods of the "cosine", "sine" and "saw"
dependencies $n_{\rm B}(Z)$,
Eq. (\ref{CaseStudies:eq.01a})-(\ref{CaseStudies:eq.01c}).
}
\label{Figure02.fig}
\end{figure}

These dependencies are shown in Fig. \ref{Figure02.fig}.

Using the aforementioned dependencies in
Eqs. (\ref{Methodology:eq.09a})-(\ref{Methodology:eq.09})
one writes the profile (\ref{Methodology:eq.12}) as
follows:
\begin{eqnarray}
y(z)
=
\textcolor{blue}{-}
{\kappa \over 4}
\left[
 \left(n_{\rm B}^{\max} + n_{\rm B}^{\min}\right)  z
-\left(n_{\rm B}^{\max} - n_{\rm B}^{\min}\right) f(z)
\right]
\label{CaseStudies_new:eq.10}
\end{eqnarray}
where $f(z)$ is a periodic function.
Its explicit expression reads\footnote{To avoid confusion with the
names of the trigonometric functions used in this expression
and the terms "cosine" and "sine" we note that the latter
denote the doping dependencies,
Eqs. (\ref{CaseStudies:eq.01a})-(\ref{CaseStudies:eq.01b}).}:
\begin{eqnarray}
f(z)
=
 \left\{
\begin{array}{ll}
\displaystyle
{\lamu \over 2 \pi} \sin{2\pi z \over \lamu},
& \mbox{"cosine" doping}\\
\displaystyle
-{\lamu  \over 2\pi}
\cos{2\pi z \over \lamu},
& \mbox{"sine" doping}\\
\lamu
 \left\{
\begin{array}{ll}
\displaystyle
{z \over \lamu}
- k
-
2\left({z \over \lamu} - k\right)^2\!\! ,
& \displaystyle
{z \over \lamu}  = [k, k+0.5] \lamu,   \\
\displaystyle
{z \over \lamu}
- k - 1
+ 2\left(k + 1 - {z \over \lamu}\right)^2\!\!,
& \displaystyle
{z \over \lamu} = \left[k+0.5, k+1\right]
\end{array}
\right.
,
& \mbox{"saw" doping}
\end{array}
\right.
\label{CaseStudies_new:eq.11}
\end{eqnarray}
The term $\kappa n_0 f(z)/4$ on the right-hand side of
Eq. (\ref{CaseStudies_new:eq.10}) represents the undulatory part of
the profile.
Its period is $\lamu = \sqrt{2} \lambda_{\rm B}$ and the
amplitude value $a$ is
$\kappa\left(n_{\rm B}^{\max} - n_{\rm B}^{\min}\right)\lamu/8\pi$
for the "sine" and "cosine" doping and to a smaller value of
$\kappa\left(n_{\rm B}^{\max} - n_{\rm B}^{\min}\right)\lamu/32$ for the "saw" doping (see also
Fig. \ref{Figure09.fig} in Appendix \ref{Profiles} which compares
the profiles of $f(z)/\lamu$).
For boron-doped diamonds produced by Microwave Plasma Chemical
Vapor Deposition (MPCVD), typical values of
$\lamu$ are in the range $5-20$ microns and
$n_{\rm B}^{\max} - n_{\rm B}^{\min}\sim 10^{21}$ cm$^{-3}$
\cite{ThuNhi_Private}.
These values together with $\kappa$ from Eq.
(\ref{K_coefficient:eq.02}) lead to an estimate of
$a < 10$ \AA.

The term proportional to $z$ on the right-hand side of
Eq. (\ref{CaseStudies_new:eq.10}) determines the direction
of the centreline of the undulating profile
(i.e., the undulator axis, labelled as $z^{\prime}$ in Fig.
\ref{Figure01.fig}) with respect to the (-110) plane in the
substrate.
Its derivative gives the angle
$\alpha_1 \ll 1$ between the plane and the centreline:
\begin{eqnarray}
\alpha_1
=
{\kappa  \left(n_{\rm B}^{\max} + n_{\rm B}^{\min}\right) \over 4}\,.
\label{CaseStudies_new:eq.12}
\end{eqnarray}
The value of $\alpha_1$ can be in the range
of several hundred microradians for boron-doped diamonds.

As mentioned in Section \ref{Introduction}, the PB boron-doped
segment of the hetero-crystal cannot be separated from the substrate.
As a rule, the thickness of the substrate in the $Z$ direction
(hundred of microns and more)
is notably larger than the thickness of the doped layer (few tens of
microns).
Therefore, in the experiment, in order to increase the intensity of the
undulator radiation, the beam of particles must be incident on the
PB segment to avoid the decrease in the number of channeling particles
due to the dechanneling in the substrate
\cite{PavlovEtAl:SPB_v14_190_2021}.
In this case, to increase the acceptance of the beam particles into the
channeling mode, the angle between the beam direction and a
tangent to the PB profile at the crystal entrance must be much
smaller than Lindhard's
critical angle $\theta_{\rm L}$ \cite{Lindhard}.
Ideally, the beam must be aligned with the tangent line, the angle
$\alpha \ll 1$ of which with respect to the $z$-axis is
\begin{eqnarray}
\alpha
=
\left.{\d y(z) \over \d z}\right|_{z=0}
=
\alpha_1
-
{\kappa \left(n_{\rm B}^{\max} - n_{\rm B}^{\min}\right) \over 4}
\left.{\d f(z) \over \d z}\right|_{z=0}\,.
\label{CaseStudies_new:eq.13}
\end{eqnarray}
Note that $(\d f(z)/\d z)_{z=0}=1$ for the
for the "cosine" and the "saw" doping
while for the "sine" doping $(\d f(z)/\d z)_{z=0}=0$.
The latter case may be preferable for channeling experiments
as it requires the beam to be aligned with the centreline, the
direction of which can be precisely defined.

The derivative $\d y(z)/\d z$ calculated at $z=L_{\rm u}$
determines the direction of the tangent line at the end of the
boron-doped layer.
This quantity determines the mean angle $\langle \alpha \rangle$
of the channeling particles at the entrance to the substrate.
By a proper choice of the dopant concentration in the vicinity of
$z=L_{\rm u}$ it is possible to fulfill the
$\langle \alpha \rangle \ll \theta_{\rm L}$ condition,
thereby increasing the number
of particles accepted into the channeling regime in the substrate.

The general formulae
(\ref{Methodology:eq.03})-(\ref{Methodology:eq.12})
allow the bending profile to be evaluated for any
dependence of the boron concentration on the $Z$ coordinate.
In particular, the dependence $n_{\rm B}(Z)$,
used in the process of growing boron-doped diamonds and
measured by the secondary-ion mass spectrometry (SIMS) method
\cite{BackeEtAl:arXiv_2404.15376v3},
can serve as input data to calculate $y(z)$.
To illustrate this, in Section \ref{CaseStudy_2}
we model a quasi-periodic doping which resembles
the doping profile in the periodically bent sample grown
using the MPCVD method \cite{ThuNhi_Private}.

\section{Numerical results  \label{CaseStudies}}

\subsection{Case study I  \label{CaseStudy_1}}

In this section we present the results of numerical simulations
of the channeling process and the radiation emission by $\E=855$ MeV
electrons and 530 MeV positrons incident on the oriented diamond
crystal periodically doped with boron.
The parameters of the "saw"-doped layer used in the simulations
correspond to those described in Ref. \cite{BackeEtAl:arXiv_2404.15376v3}
and are as follows:
(i) the doping period in the [100] direction
$\lambda_{\rm B} = 5/\sqrt{2}$ $\mu$m,
(ii) the number of periods $N=4$,
(iii) the minimum and maximum values of the boron concentrations used
in the process of producing the layer by MPCVD are
$n_{\rm B}^{\min} = 6\times10^{20}$ cm$^{-3}$ and
$n_{\rm B}^{\max} = 24\times10^{20}$ cm$^{-3}$.

In the cited paper the coefficient $\kappa$, which enters into
the  dependence (\ref{Methodology:eq.05}) of the diamond lattice constant
on the boron concentration, was calculated according to Ref.
\cite{BrazhkinEtAl-PRB_v74_140502_2006}.
Since the available experimental and theoretical data on $\kappa$
(see, e.g., Fig. 2 in \cite{BackeEtAl:arXiv_2404.15376v3})
do not allow one a definite choice of its value,
the current simulations were performed for both values of
$\kappa$ given in Eq. (\ref{K_coefficient:eq.02}).

\subsubsection{\textbf{Description of the profile}
\label{CNRS+MAMI:01}}

\begin{figure} [h]
\centering
\includegraphics[clip,width=13cm]{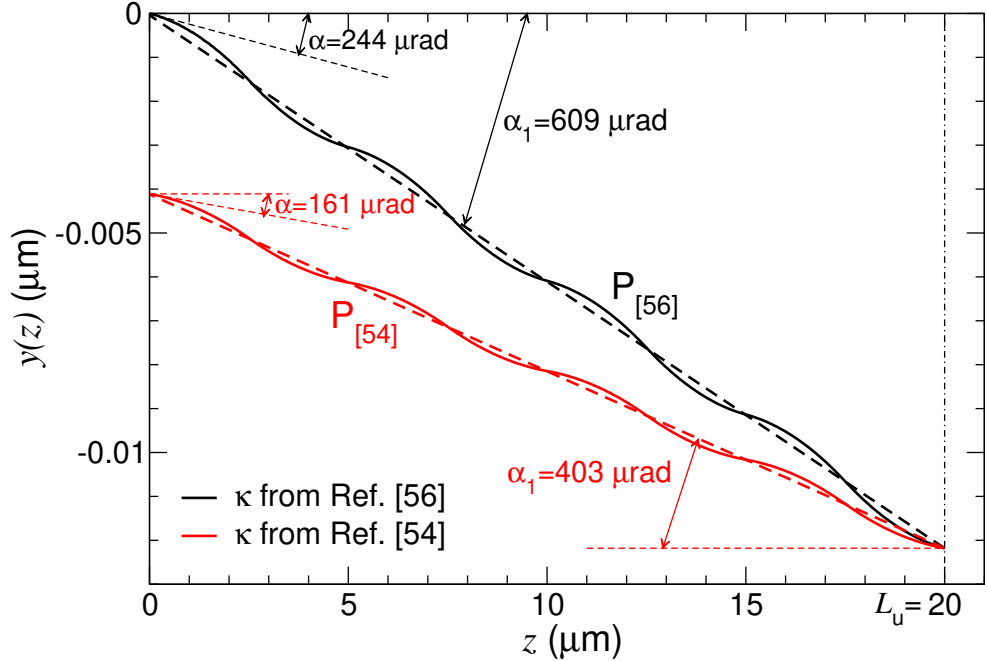}
\caption{
Solid curves show the profile $y(z)$, Eq. (\ref{CaseStudies_new:eq.10}),
dashed lines show the centreline of the profile.
Two sets of dependencies correspond to the values of $\kappa$
from Refs. \cite{BrazhkinEtAl-PRB_v74_140502_2006} and \cite{Vegard:ZPhysik_v5_p17_1921}.
The corresponding profiles are designated as
P$_{\mbox{\footnotesize \cite{BrazhkinEtAl-PRB_v74_140502_2006}}}$
and
P$_{\mbox{\footnotesize \cite{Vegard:ZPhysik_v5_p17_1921}}}$.
Vertical dash-dotted line marks the thickness $L_{\rm u}=20$ $\mu$m
of the doped layer along the (-110) planar direction,
the $z$ axis.
Also shown for each profile are the angles
$\alpha_1$, Eq. (\ref{CaseStudies_new:eq.12}),
and
$\alpha$, Eq. (\ref{CaseStudies_new:eq.13}),
between the $z$ axis and
the centreline and the tangent line at $z=0$, respectively.
}
\label{Figure03.fig}
\end{figure}

Figure \ref{Figure03.fig} shows the profiles $y(z)$ (solid lines)
along the (-110) planar direction calculated for
$\kappa_{\mbox{\footnotesize \cite{BrazhkinEtAl-PRB_v74_140502_2006}}}$
and
$\kappa_{\mbox{\footnotesize \cite{Vegard:ZPhysik_v5_p17_1921}}}$.
For brevity, the notations
P$_{\mbox{\footnotesize \cite{BrazhkinEtAl-PRB_v74_140502_2006}}}$
and
P$_{\mbox{\footnotesize \cite{Vegard:ZPhysik_v5_p17_1921}}}$
are used to distinguish the profiles.
The dashed lines represent the centrelines; the corresponding values of
$\alpha_1$, Eq. (\ref{CaseStudies_new:eq.12}), are indicated
explicitly.
The amplitudes of the undulatory parts of the profiles are
$a_{\mbox{\footnotesize \cite{Vegard:ZPhysik_v5_p17_1921}}}=2.3$ \AA{} and
$a_{\mbox{\footnotesize \cite{BrazhkinEtAl-PRB_v74_140502_2006}}}=1.5$ \AA{}.
The period $\lamu = 5$ $\mu$m is the same for each
profile.
The vertical line marks the thickness of the doped layer,
$L_{\rm u}=4\lamu =20$ $\mu$m.
In the point $z=0$ the absolute values of the angle
(\ref{CaseStudies_new:eq.13})
between the tangent lines and the $z$ axis are
\begin{eqnarray}
\left|\alpha_{\mbox{\footnotesize \cite{Vegard:ZPhysik_v5_p17_1921}}}\right|
=
244\, \mbox{$\mu$rad},
\qquad
\left|\alpha_{\mbox{\footnotesize \cite{BrazhkinEtAl-PRB_v74_140502_2006}}}\right|
=
161\, \mbox{$\mu$rad}.
\label{CNRS+MAMI:eq.10}
\end{eqnarray}

Apart from the quantities mentioned above, it is instructive to
calculate the curvature $R^{-1}=|y^{\prime\prime}|/(1 + y^{\prime\,2})^{3/2}$
of the profiles.
In a bent crystal, this quantity defines the centrifugal force
$F_{\rm cf} \approx \E/R$ acting on the channeling particle in
the co-moving frame.
The condition for a stable channeling \cite{Tsyganov1976} implies that
$F_{\rm cf}$ is less than the maximum interplanar force
$U^{\prime}_{\max}$, which holds the particle in the channel.
Using Eqs. (\ref{CaseStudies_new:eq.10}) and
(\ref{CaseStudies_new:eq.11}), one derives the general expression
$R^{-1}\approx|y^{\prime\prime}|
= \kappa \left(n_{\rm B}^{\max} - n_{\rm B}^{\min}\right)/\lamu$ and,
as a consequence, the values
$R^{-1}_{\mbox{\footnotesize \cite{Vegard:ZPhysik_v5_p17_1921}}}=2.94$ cm$^{-1}$ and
$R^{-1}_{\mbox{\footnotesize \cite{BrazhkinEtAl-PRB_v74_140502_2006}}}=1.94$ cm$^{-1}$.
In the framework of the continuous potential model \cite{Lindhard},
the maximum interplanar force in the diamond (110) channel
is estimated to be $U^{\prime}_{\max}=7.8$ GeV/cm.
Then, the values of the bending parameter $C$ defined as the ratio
$F_{\rm cf}/U^{\prime}_{\max}$ are as follows:
\begin{eqnarray}
\begin{array}{ll}
C_{\mbox{\footnotesize \cite{Vegard:ZPhysik_v5_p17_1921}}} = 0.20,
\
C_{\mbox{\footnotesize \cite{BrazhkinEtAl-PRB_v74_140502_2006}}} = 0.11
& \mbox{for $\E$=530 MeV}, \\
C_{\mbox{\footnotesize \cite{Vegard:ZPhysik_v5_p17_1921}}} = 0.32,
\
C_{\mbox{\footnotesize \cite{BrazhkinEtAl-PRB_v74_140502_2006}}} = 0.22
& \mbox{for $\E$=855 MeV}.
\end{array}
\label{Curvature:eq.08}
\end{eqnarray}

Note that for both profiles the curvature radius is constant.
Therefore, the periodic component of each profile is a series of
arcs of circles of equal radius but alternating sign of the curvature.

\subsubsection{\textbf{Results of atomistic simulations of the
channeling and photon emission processes}
\label{Initial_v}}

For each profile, electron and positron passage simulations and
photon emission spectra calculations were performed for the
following two different beam–crystal alignments (\textbf{A}):

\begin{itemize}

 \item[\textbf{A$_1$:}]
  The incident beam velocity $\bfv_0$
  is aligned with the (-110) planar direction
  in the substrate (i.e., parallel to the
 $z$ axis in Figs. \ref{Figure01.fig} and  \ref{Figure03.fig}).

 It should be noted that this alignment was used in Ref. \cite{BackeEtAl:arXiv_2404.15376v3} (see Sections 5.1 and 5.2 there).

 \item[\textbf{A$_2$:}]
  The velocity is aligned with the tangent line to the profile
  at $z=0$.
\end{itemize}

%
The root-mean square divergences $\phi_{x,y}$ of the electron and positron
Gaussian beams used in the simulations
were calculated using the full width at half maximum data from
Refs. \cite{BackeEtAl:EPJD_v76_150_2022,Backe:NIMA_v1059_168998_2024}:
$(\phi_{x},\phi_{y})=(80,10)$ and (270,27) $\mu$rad for the electron
and positron beams, respectively.
To ensure statistical reliability of the results, in each simulation
(i.e., for each projectile type, each profile and alignment)
approximately $12\times 10^3$ trajectories were simulated.

\begin{figure} [h]
\centering
\includegraphics[clip,width=13cm]{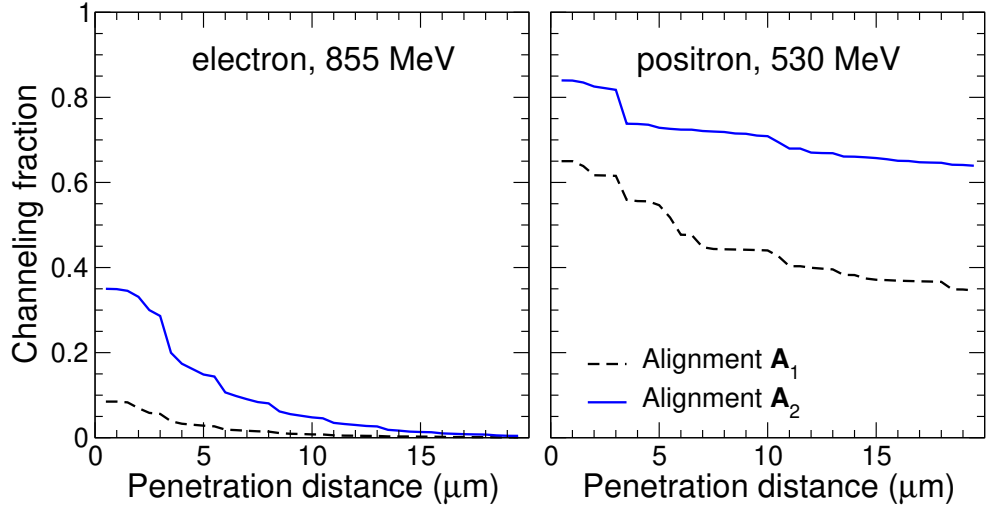}
\caption{
Channeling fractions of accepted particles versus penetration
distance calculated for 855 MeV electrons (left) and 530 positrons
(right).
Dashed and solid curves correspond to the two alignments as indicated
in the common legend.
The results shown refer to the profile with
$\kappa=\kappa_{\mbox{\footnotesize \cite{BrazhkinEtAl-PRB_v74_140502_2006}}}$,
see Eq. (\ref{K_coefficient:eq.02}).
}
\label{Figure04.fig}
\end{figure}

The impact of the choice of the beam–crystal orientation on the channeling
efficiency is illustrated in Fig. \ref{Figure04.fig}, which compares the
dependence of the fraction of particles, which were accepted in the
channeling mode at the entrance, on the penetration distance.
The left and right panels refer to the electrons and positrons, respectively.
In both panels, the dashed curves show the dependence obtained for the
alignment \textbf{A}$_1$, the solid curves -- for \textbf{A}$_2$.
All dependencies shown refer to the profile
P$_{\mbox{\footnotesize \cite{BrazhkinEtAl-PRB_v74_140502_2006}}}$.

In the case of \textbf{A}$_2$, the fractions $\calA$ of the particles
accepted at the entrance are about 0.4 and 0.8 for the electrons and positrons,
respectively.
These values correlate with the previously calculated data for a straight
diamond (110) channel: $\calA_0\approx 0.7$ for 855 MeV electrons
\cite{Pavlov-EtAl:EPJD_2020,SushkoEtAl:arXiv_2404.15376v3}
and $\calA_0\approx 0.95$ for 530 MeV positrons
\cite{SushkoEtAl:arXiv_2404.15376v3}.
To see the correspondence, recall that for an ideally collimated beam
the acceptances in the bent and straight channels are related as
$\calA=(1-C)\calA_0$ (see, e.g., \cite{BiryukovChesnokovKotovBook})
and use the values $C_{\mbox{\footnotesize \cite{BrazhkinEtAl-PRB_v74_140502_2006}}}$ given in (\ref{Curvature:eq.08}).
The remaining discrepancy could be attributed to the fact that the simulations \cite{Pavlov-EtAl:EPJD_2020}
were performed for the beams with non-zero divergence.

Comparing the dashed and the solid curves in Fig. \ref{Figure04.fig}, it can
be  seen that the fractions are much smaller for the \textbf{A}$_1$
orientation.
The decrease is much more pronounced for electrons.
This can be understood comparing the angle $\alpha = 161$ $\mu$rad
with Lindhard's critical angle $\theta_{\rm L}(C)$ in the bent diamond
(110) channel.
In the straight channel, $\theta_{\rm L}(0)=\sqrt{2U_0 /\E}$ where
$U_0\approx 20$ eV is the depth of the interplanar potential.
In the bent channel, this value is multiplied by a factor $(1-C)$
(see, e.g., \cite{Taratin_PhysPartNucl_v29_p437_1998-English}).
Using the values $C_{\mbox{\footnotesize \cite{BrazhkinEtAl-PRB_v74_140502_2006}}}$, one finds
$\theta_{\rm L}(C) \approx 164$ and $\approx 240$ $\mu$rad for the electrons
and positrons, respectively.
This estimate shows that for the electrons
$\alpha \approx \theta_{\rm L}(C)$ and, therefore, a small fraction
of the beam incident along the $z$ axis is accepted.
In contrast, the inequality $\alpha < \theta_{\rm L}(C)$
obtained for positrons significantly increases the probability of being captured in the channeling regime.

A similar analysis performed for the profile
P$_{\mbox{\footnotesize \cite{Vegard:ZPhysik_v5_p17_1921}}}$
shows that the fractions for the \textbf{A}$_2$ orientation are slightly
reduced due to the larger values of $C$ (see Eq. (\ref{Curvature:eq.08})).
In contrast, the fraction of channeling particles
for the \textbf{A}$_1$ orientation virtually disappears for both types of
projectiles.
The estimates show that in this case the angle $\alpha$ exceeds
$\theta_{\rm L}(C)$ leading to a strong reduction in the number of accepted
particles.

\begin{figure} [h]
\centering
\includegraphics[clip,width=15cm]{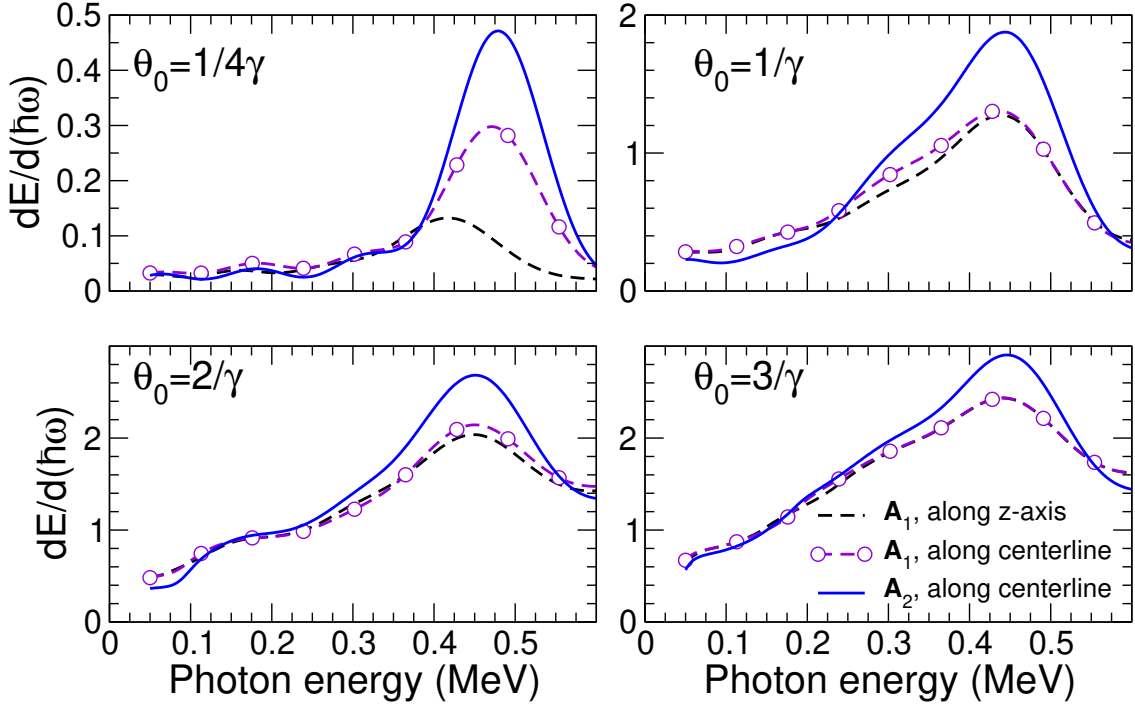}
\caption{
Spectral distributions of radiation for 530 MeV positrons
for the profile
P$_{\mbox{\footnotesize \cite{BrazhkinEtAl-PRB_v74_140502_2006}}}$,
see Fig. \ref{Figure03.fig}.
Each graph corresponds to the emission within a given emission
cone $\theta_0$ given in units of $1/\gamma\approx 960$ $\mu$rad.
Dashed curves without symbols show the spectra calculated for
the alignment \textbf{A$_1$} with the emission cone along the incident beam
velocity $\bfv_0$ ("z-axis").
Dashed curves with symbols correspond to \textbf{A$_1$} but with the
emission cone directed along the centreline of the profile, i.e. at
403 $\mu$rad to $\bfv_0$.
Solid curves correspond to \textbf{A$_2$}
with the emission cone along the centreline.
Note that the values of $\d E/\d(\hbar\om)$ are multiplied by a factor
of $10^3$.
}
\label{Figure05.fig}
\end{figure}

\begin{figure} [h]
\centering
\includegraphics[clip,width=15cm]{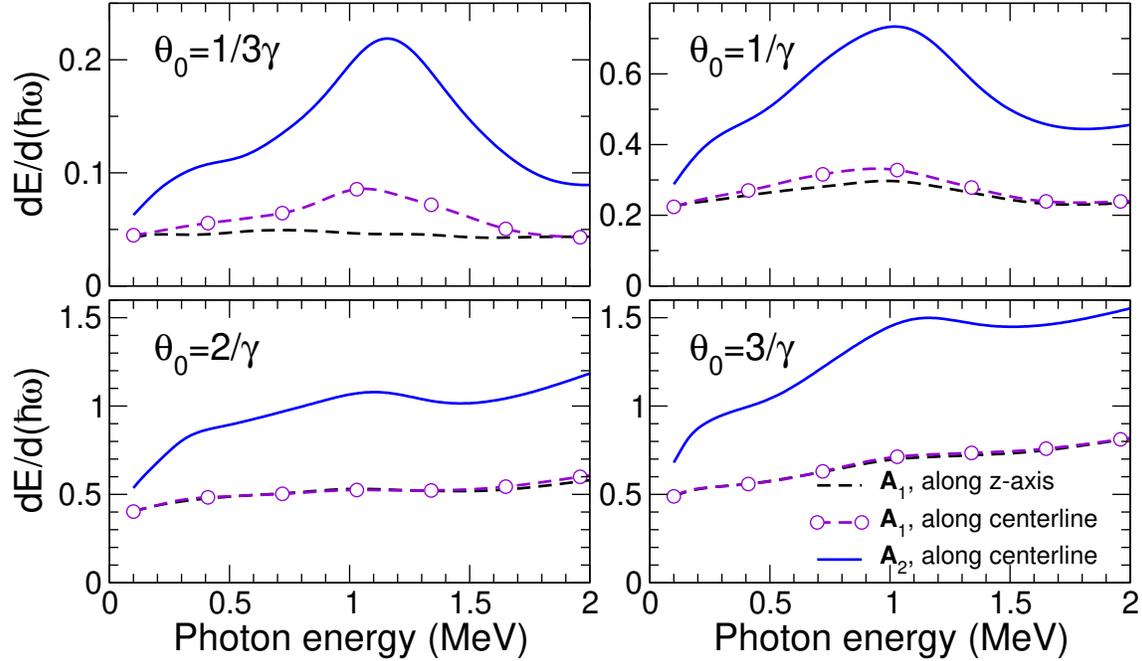}
\caption{
Same as in Fig. \ref{Figure05.fig} but for 855 MeV electrons.
}
\label{Figure06.fig}
\end{figure}

\begin{figure} [h]
\centering
\includegraphics[clip,width=15cm]{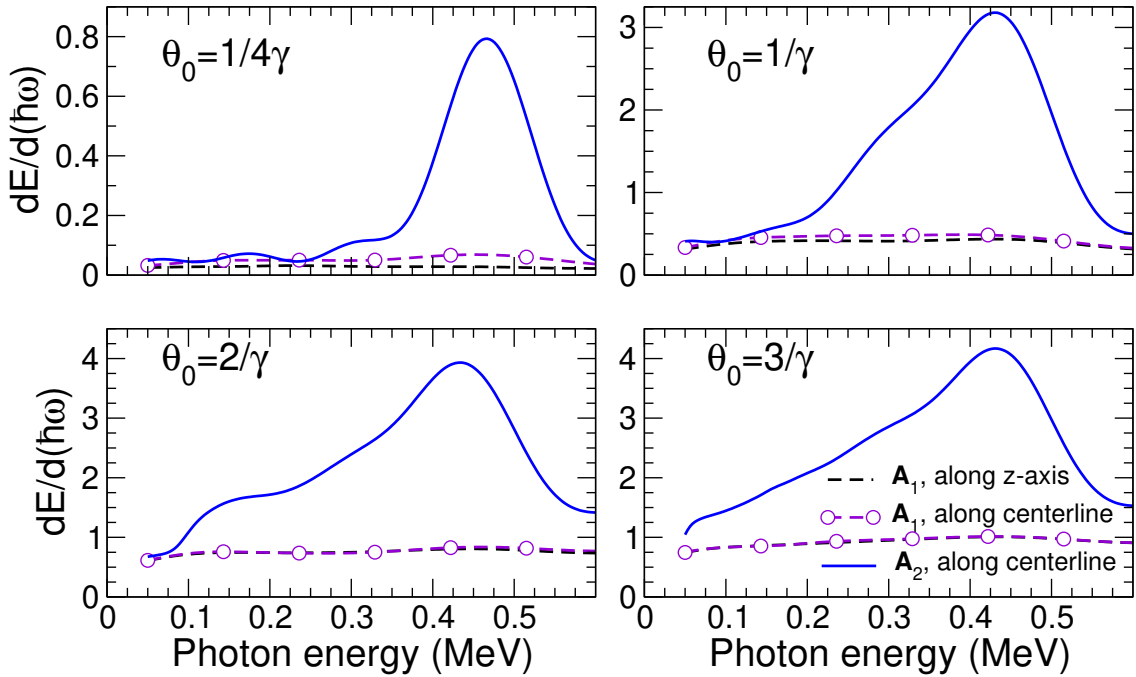}
\caption{
Same as in Fig. \ref{Figure05.fig} but for the
P$_{\mbox{\footnotesize \cite{Vegard:ZPhysik_v5_p17_1921}}}$ profile.
}
\label{Figure07.fig}
\end{figure}

\begin{figure} [h]
\centering
\includegraphics[clip,width=15cm]{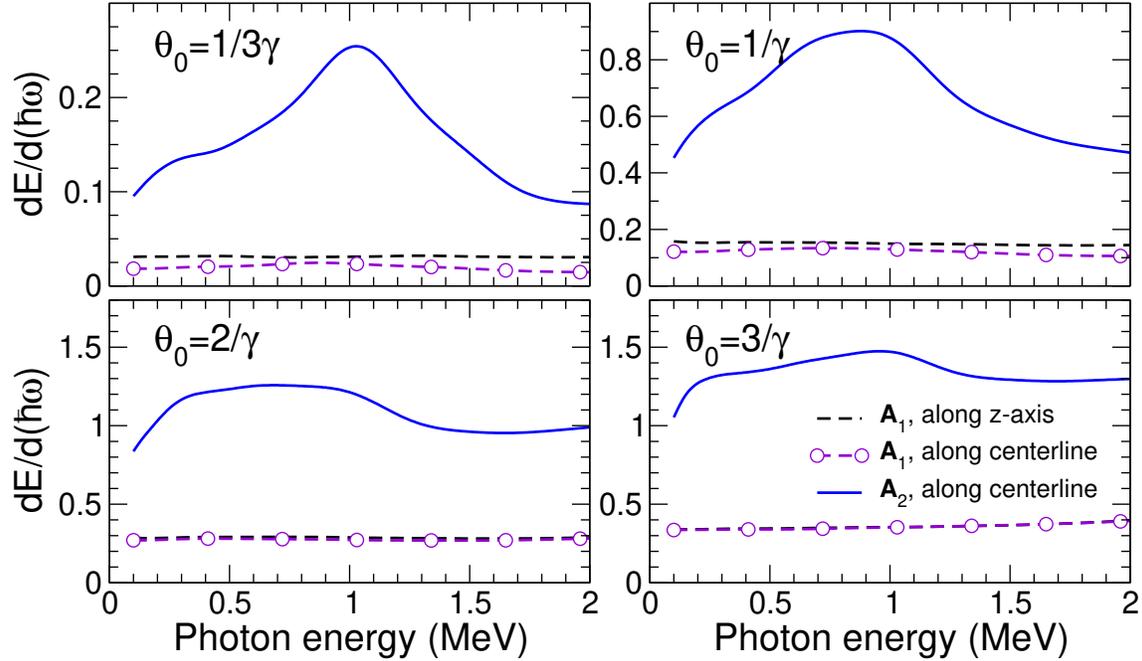}
\caption{
Same as in Fig. \ref{Figure06.fig} but for the
P$_{\mbox{\footnotesize \cite{Vegard:ZPhysik_v5_p17_1921}}}$ profile.
}
\label{Figure08.fig}
\end{figure}

The differences in channeling efficiency for the two orientations and for the
two profiles are clearly visible in the spectral distributions of the
emitted radiation, shown in Figs. (\ref{Figure05.fig})-(\ref{Figure06.fig})
for the profile P$_{\mbox{\footnotesize \cite{BrazhkinEtAl-PRB_v74_140502_2006}}}$ and in Figs. (\ref{Figure07.fig})-(\ref{Figure08.fig})
for P$_{\mbox{\footnotesize \cite{Vegard:ZPhysik_v5_p17_1921}}}$.

A particle channeled in a periodically bent crystal,
experiences two types of quasi-periodic motion:
the channeling oscillations and the crystal undulator (CU) oscillations
due to the periodicity of the  bent profile.
These motions are very similar to the undulatory motion.
As a result, for each value of the emission angle $\theta$ the constructive interference of the waves occurs at some particular equally spaced
frequencies (harmonics), so that the spectral
distribution consists of a series of narrow peaks.
As a rule, the characteristic frequency of the channeling oscillations exceeds greatly that of the CU ones and, therefore, the peak(s) of the
channeling radiation \cite{ChRad:Kumakhov1976} are located at much higher photon
energies than the peaks of CU radiation (CUR) \cite{ChannelingBook2014}.

The spectral distributions $\d E/\d(\hbar\om)$ of radiation shown in
Figs. (\ref{Figure05.fig})-(\ref{Figure08.fig}) refer to
the photon energy range where CUR dominates.

For each simulated trajectory, the spectral-angular distribution
of radiation $\d E^3/\d(\hbar\om) \d \Om$ emitted within
the solid angle $\d\Om\approx \theta \d\theta\d\phi$
(where $\theta$ and $\phi$ are the polar angles of the emission
direction) is calculated numerically following the algorithm
outlined in Refs. \cite{MBN_ChannelingPaper_2013,ChannelingBook2014}.
The spectral distribution of energy radiated within the cone
$\theta_0\ll1$ along {\em any chosen} direction and averaged over
all trajectories is calculated as follows:
\begin{eqnarray}
{\d E(\theta\leq\theta_{0}) \over \d (\hbar\om)}
=
{1 \over N}
\sum_{n=1}^{N}
\int\limits_{0}^{2\pi}
\d \phi
\int\limits_{0}^{\theta_{0}}
\theta \d\theta\,
{\d^3 E_n \over \d (\hbar\om)\, \d\Om}
\label{Spectra:eq.01}
\end{eqnarray}
where summation is made over all simulated trajectories
of the total number $N$.

The structure of each of the figures is the same and it is as follows.
The four graphs correspond to different emission cones (as indicated)
measured in units of $1/\gamma$ equal to approximately 0.96 and 0.60
mrad for the 530 and 855 MeV projectiles, respectively.
The cones $\theta_0=1/4\gamma$ for $\E=530$ MeV and
$\theta_0=1/3\gamma$ for 855 MeV shown in the upper left graphs
are close to the value of $0.24$ mrad used in
Ref. \cite{BackeEtAl:arXiv_2404.15376v3}.
The dashed curves refer to the spectra calculated for the \textbf{A$_1$}
beam-crystal alignment: (i) the ones without symbols stand for the
emission cone is chosen along the incident beam velocity,
(ii) the curves with open circles show the
intensity of the radiation, which is emitted in the cone
along the profile centreline.
The solid curves refer to the radiation emission along the centreline
calculated for the \textbf{A$_2$} alignment.

The common feature of both types of the projectile, both profiles and
all emission cones, is that the most intense radiation is emitted when the beam impinges the crystal along the tangent to the profile.
This is not surprising since the $\textbf{A$_2$}$  orientation maximises the number of
channeling particles emitting CUR, whose powerful first harmonic
maxima are seen in the photon energy range 0.4-0.5 MeV
for the positrons and 0.5-1.5 MeV for the electrons.

The energy $\hbar\om_1$ of the fundamental harmonic of the
undulator radiation can be calculated as follows (see, e.g., \cite{Elleaume:RevSciInstrum_v63_p321_1991,CLS-book_2022}):
\begin{eqnarray}
\hbar\om_1\, \mbox{[MeV]}
\approx
{ 9.5 \E^2 \, \mbox{[GeV]} \over \lambda_{\rm u}\, \mbox{[$\mu$m]}}
{1 \over 1 + K^2/2 + (\gamma\theta)^2 }
\label{MBN_Emission:eq.03}
\end{eqnarray}
The undulator parameter $K$ is related to the mean square
velocity of the particle in the transverse direction:
$K^2 = 2\gamma^2 \langle v_{\perp}^2 \rangle /c^2$.
In a CU this parameter accounts for both the undulator oscillations and
the channeling oscillations \cite{Dechan01}:
\begin{eqnarray}
K^2 = K_{\rm u}^2 + K_{\rm ch}^2\,.
 \label{MBN_Emission:eq.02}
\end{eqnarray}
It can be shown that for an ultra-relativistic particle
moving along the periodically bent channel
(\ref{CaseStudies_new:eq.10})
$K_{\mathrm{u}}=\xi \gamma a/\lamu$ with $\xi=2\pi$ for the
"cosine" and "sine" doping, and $\xi=8\sqrt{2/3}$ for the "saw"
doping.
For positron channeling, assuming the channeling
oscillations are harmonic, one relates the value of
$K_{\mathrm{ch}}^2$  averaged over the amplitude of channeling
oscillations to the depth of the interplanar potential:
$\langle K_{\rm ch}^2 \rangle  = {2\gamma U_0 / 3mc^2}$
(see, e.g., \cite{ChannelingBook2014}, Eq. (B.5)).

For the smallest values of the emission cones
$\theta_0 = 1/3\gamma$ and $1/4\gamma$, the
positions of the peaks are in a very good agreement with the
values calculated from Eq. (\ref{MBN_Emission:eq.03}).
Increase in $\theta_0$ results in the enhancement of the radiation
in the off-axis directions, which is emitted at lower photon
energies energies.
As a result, the peak become broader and its position is
red-shifted.

It can be seen that for the $\textbf{A$_1$}$ beam-crystal
alignment the calculated spectra are very sensitive to
the value of the coefficient $\kappa$, Eqs.
(\ref{Methodology:eq.05})-(\ref{K_coefficient:eq.02}),
chosen to model the bending profile.
As mentioned above, for
$\kappa=\kappa_{\mbox{\footnotesize \cite{Vegard:ZPhysik_v5_p17_1921}}}$ the fraction of the
particles accepted in the channeling mode is negligibly
small.
Therefore, in this case almost all beam particles
experience the over-barrier motion and instead of CUR they
emit the incoherent bremsstrahlung radiation, which
produces a smooth background in the spectral distribution.
This is clearly seen in Figs.
(\ref{Figure07.fig})-(\ref{Figure08.fig}) for both
electrons and positrons and for all emission cones.

For the smaller value of the coefficient,
$\kappa=\kappa_{\mbox{\footnotesize \cite{BrazhkinEtAl-PRB_v74_140502_2006}}}$, the spectral distributions of
the radiation emitted in the cones along the centreline
(dashed lines with symbols) are, basically, similar in shape to those obtained for the $\textbf{A$_2$}$
alignment but less intense due to the smaller number
of the particles that channel in the
periodically bent channel.
The distributions shown in the dashed curves without
symbols represent the radiation from the channeling
particles but emitted off-axis, i.e. not along the
centreline, which is the undulator axis,
but in the cone whose axis is inclined at
the angle $\alpha_1=403$ $\mu$rad to
the centreline.
From the general theory of the undulator radiation (see, e.g., \cite{AlferovBashmakovCherenkov1989}) it is known that
the off-axis radiation is less intense than the
on-axis radiation and that its characteristic energy is red-shifted.
The latter property follows directly from
Eq. (\ref{MBN_Emission:eq.03}) where the emission angle
$\theta$ must be considered to be centered at
$\alpha_1$.
Therefore, the term $(\gamma\alpha_1)^2$ in the
denominator reduces the energy $\hbar\om_1$.
The shift in the peak position can be clearly seen for
in the spectral distribution of radiation emitted by positrons in the small emission cone $1/4\gamma$,
Fig. \ref{Figure05.fig}.
In the electron spectrum, top left graph in Fig. \ref{Figure06.fig}, it is much less pronounced, but still
distinguishable as a hump around
$\hbar\om =0.7$ MeV.
For larger emission cones, at
$\theta_0 \gg \alpha_1$, the difference between
the on-axis and off-axis spectral distributions virtually
disappears because these cones collect all the emitted radiation.

\subsection{Case study II  \label{CaseStudy_2}}


In this section we present the results of numerical simulations
of the channeling process and the radiation emission by $\E=855$ MeV
electrons and 530 MeV positrons in a diamond
crystal doped quasi-periodically.

We consider a quasi-periodic doping, which models the doping profile
in the sample grown using the MPCVD method \cite{ThuNhi_Private}.
\begin{figure} [h]
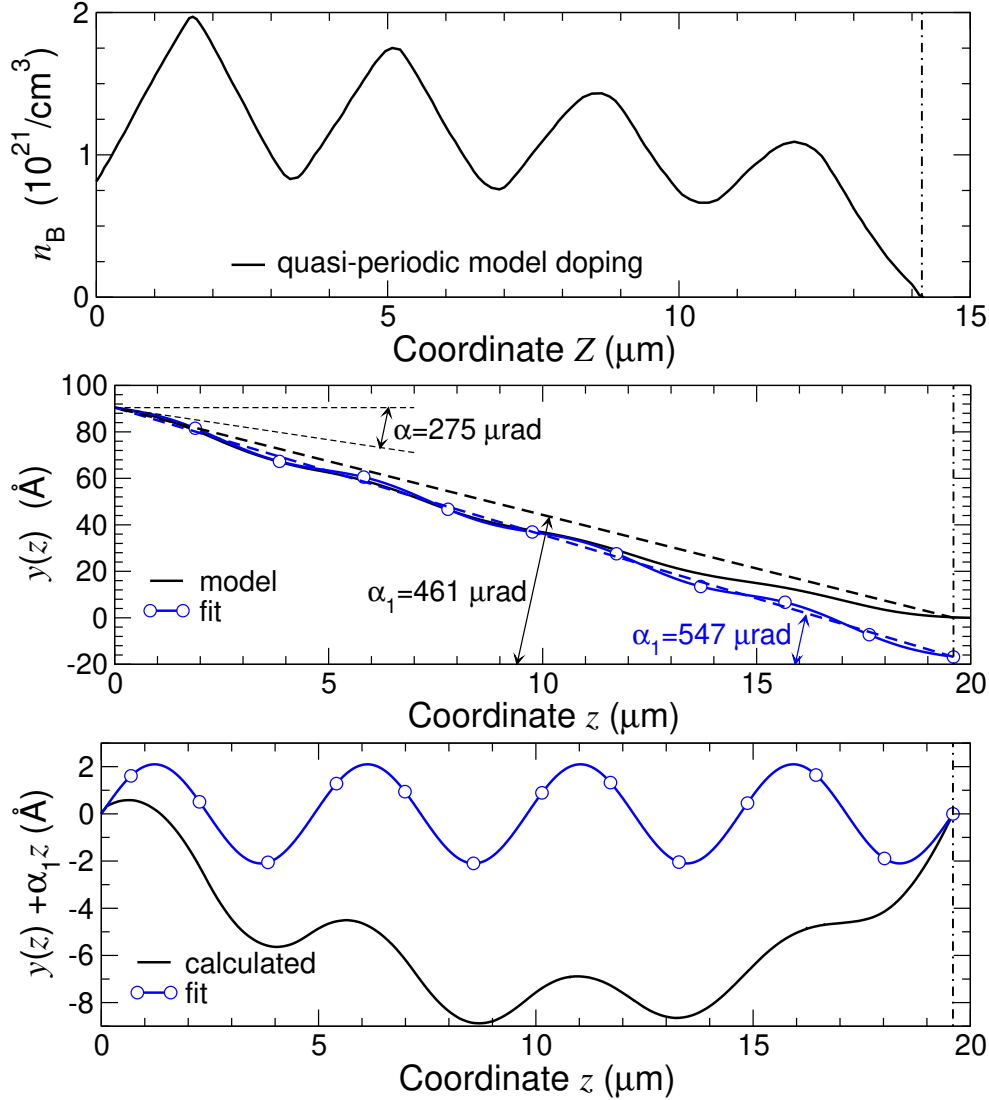

\centering
\includegraphics[clip,width=13cm]{Figure10a.eps}
\includegraphics[clip,width=13cm]{Figure10b.eps}
\includegraphics[clip,width=13cm]{Figure10c.eps}
\caption{
Four-period boron-doped diamond with model doping.
\textit{Top:} Model boron concentration $n_{\rm B}(Z)$ versus
the coordinate along the [100] axis in ca 14 $\mu$m thick
doped layer.
\textit{Middle:} The model profile of periodically bent (-110) planes
calculated for $n_{\rm B}(Z)$ (black solid curve)
and its analytical fit with sinusoidal function
(blue solid curve with open circles).
The profiles are shown in the $(y,z)$ coordinates, see
Fig. \ref{Figure01.fig}.
For the fit, the dashed line shows the profile's centreline,
which is inclined to the $z$ axis by $\alpha_1=547$ $\mu$rad.
For the calculated profile, the dashed line, inclined by
$461$ $\mu$rad, connects its first, $z=0$,
and last, $z=19.6$ $\mu$m, points.
For both profiles, the angle between the tangent line in the first
point and the $z$ axis is 275 $\mu$rad.
The profiles correspond to
$\kappa=\kappa_{\mbox{\footnotesize\cite{Vegard:ZPhysik_v5_p17_1921}}}$.
\textit{Bottom:} Quasi-periodic (for the calculated profile)
and periodic (for the fit) parts of the profiles.
These curves are obtained by subtracting the linear dependencies
(dashed lines) from the solid curves in the middle graph.
\textit{In all graphs}, the vertical dash-dotted line marks the end of the doped layer.
}
\label{Figure10.fig}
\end{figure}

The boron concentration $n_{\rm B}(Z)$ versus distance $Z$
from the substrate along the [100] direction is shown in
Fig. \ref{Figure10.fig}, top graph.
The doped layer of approximately 14 $\mu$m thickness
has four periods,
which vary in the the range 3-4 $\mu$m.
Except for the final segment of about 2 $\mu$m, the amplitude
of $n_{\rm B}(Z)$ increases nearly linearly with $Z$.
It can be seen that there are large deviations from the
 $n_{\rm B}(Z)$ dependence in the case of ideal "saw" doping,
see Fig. (\ref{Figure02.fig}).

The corresponding model profile $y(z)$ of the (-110) plane,
calculated using Eqs.
(\ref{Methodology:eq.03})-(\ref{Methodology:eq.12})
with
$\kappa=\kappa_{\mbox{\footnotesize\cite{Vegard:ZPhysik_v5_p17_1921}}}$,
is shown in the middle graph by the solid black line.
The (black) dashed line connects the first, $z=0$,
and last, $z=L_{\rm u}=19.6$ $\mu$m, points of the profile.
The solid blue curve with open circles shows the fit,
$y(z)=-\alpha_1z + a\sin(2\pi/\lamu)$, with $\alpha_1=547$ $\mu$rad
(shown in the graph), $a=2.1$ \AA, and $\lamu=4.9$ $\mu$m.
The (blue) dashed line shows the centerline of the fit.
At the entrance point $z=0$ the tangent lines to the
fitting curve and to the calculated profile are inclined by the same
angle of 275 $\mu$rad to the $z$ axis, which is aligned with the
(-110) planar direction in the substrate.

The quasi-periodic (for the model profile) and periodic (for the fit)
parts of the profiles $y(z)$ are shown in the bottom graph.
They are calculated by subtracting the linear dependencies
$-\alpha_1z$ from $y(z)$.

For the calculated profile and for its fit,
electron and positron passage simulations and
photon emission spectra calculations were performed.
The parameters of the beams used in the simulations were as
described Section \ref{Initial_v}.
The spectra were calculated for the beam–crystal alignment
\textbf{A$_2$}, i.e. directing the beam velocity with the tangent
line to the profile at $z=0$.

\begin{figure} [h]
\centering
\includegraphics[clip,width=14cm]{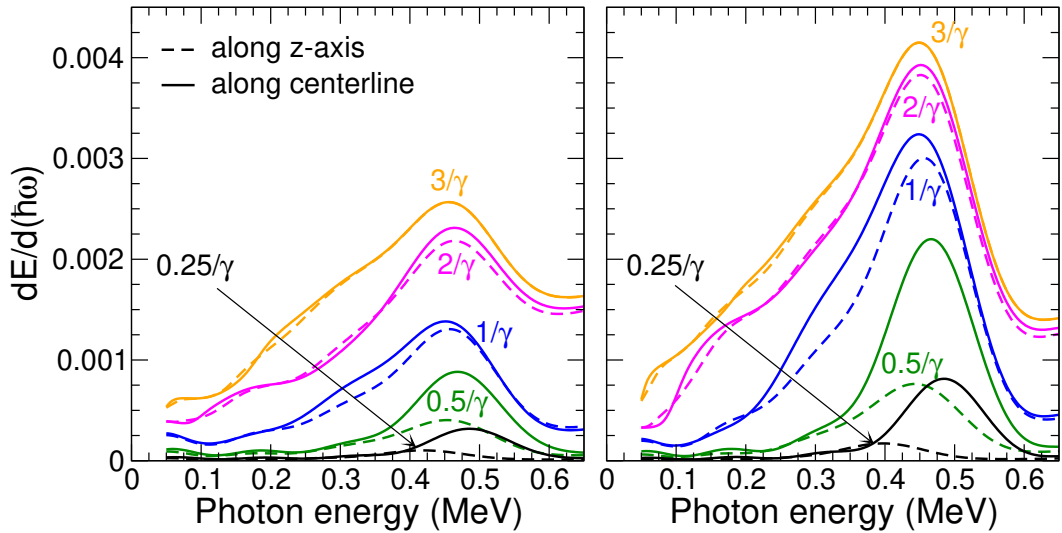}
\caption{
Spectral distributions of radiation for 530 MeV positrons
calculated for the model profile (left panel) and for its
fit (right panel).
In each graph different sets of a solid and a dashed line
corresponds to the emission within a given cone $\theta_0$
given in units of $1/\gamma\approx 960$ $\mu$rad.
Solid curves show the spectra calculated for the emission cone
along the centerline of the profile.
Dashed curves correspond to the emission cone along the $z$-axis.
}
\label{Figure11.fig}
\end{figure}

\begin{figure} [h]
\centering
\includegraphics[clip,width=14cm]{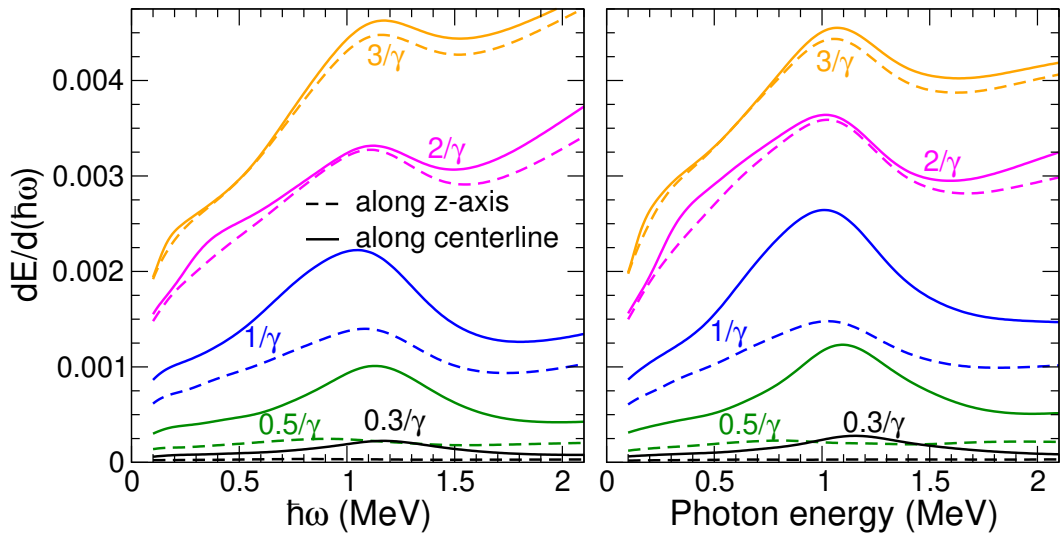}
\caption{
Same as for Fig. \ref{Figure11.fig} but for
855 MeV electrons.
The emission cones are given in units of $1/\gamma\approx 600$ $\mu$rad.
}
\label{Figure12.fig}
\end{figure}

Spectral distributions are shown in Figs.
\ref{Figure11.fig} (for 530 MeV positrons) and
\ref{Figure12.fig} (for 855 MeV electrons).
The structure of each of the figures is the same and it is as follows.
The left and right graphs represent the spectra calculated for the model
profile and for its fit, correspondingly.
The solid curves show the emission in the cone $\theta_0$
aligned with the centerline of the profile.
The dashed curves correspond to the emission along the $z$-axis.
The emission cones (as indicated) are measured in units
of $1/\gamma$ equal to approximately 0.96 and 0.60
mrad for the 530 and 855 MeV projectiles, respectively.

The feature common to both positron and electron spectra is that for the
small emission cone $\theta_0 < \alpha_1$ (see Fig. \ref{Figure10.fig},
middle graph) the emission in the forward direction (i.e. along the centerline)
is significantly more intense.
This is in line with the arguments presented in Section \ref{Initial_v}.
As discussed in the main text, the intensity of radiation emitted in
small cones is very sensitive to the direction of the cone relative to the
centerline of the profile.
For larger cones both geometries of the photon emission produce virtually the
same result.

\section{Conclusion and outlook\label{Conclusion}}

In this study, we have conducted a theoretical analysis of the
profiles of the planar (-110) crystallographic direction in
the diamond layer doped with boron atoms.
The general formalism outlined has been applied to derive the
profiles for periodic doping following the ideal dependencies
("cosine", "sine" and "saw")
of the boron concentration on the distance in the crystalline medium
as well as those measured by the
SIMS method in the samples, grown at ESRF by MPCVD method,
to be used in channeling experiments at the MAMI facility.
A detailed analysis of the geometry of the calculated profiles has been
carried out.

Numerical simulations of the channeling and photon emission processes
have been carried out for 855 MeV electron and 530 MeV positron beams
incident on boron-doped diamond with a four-period bending profile
(the bending period is $\lambda_{\rm u} = 5$ $\mu$m) corresponding
to the "saw" doping.
It is shown that the channeling efficiency and the intensity of the
crystalline undulator radiation strongly depend on the orientation of
the incident beam relative to the bent channel profile at the
entrance to the boron-doped layer.
The orientation that maximizes both mentioned quantities implies
the angle between the beam direction and the tangent line to the
channel's profile to be smaller than Lindhard's critical angle.
The direction of the tangent can be calculated unambiguously  from the
dependence of the dopant concentration, $n_{\rm B}(Z)$,  on the
coordinate along the [100] axis.
In a particular case of the "sine" doping the tangent line is along the
centreline of the periodically bent channel.
This doping scheme could be advantageous for the use in channeling
experiments, since it does not require additional alignment with the
incident beam.

In the context of a diamond hetero-crystal, it is not possible to
separate the boron-doped segment from the single diamond substrate.
Consequently, the particles propagate in the substrate after passing through the periodically bent part.
In Ref. \cite{PavlovEtAl:SPB_v14_190_2021} it was demonstrated
that in the case of positrons a comparatively large fraction of
the projectile move through the substrate in the channeling regime
and emit channeling radiation.
For electrons this fraction is negligibly small for the electrons.
The remaining beam particles undergo the overbarrier motion, thereby
emitting  incoherent bremsstrahlung.
Within the photon energy range of $\hbar\om \lesssim 1$ MeV, both
radiation mechanisms yield a smooth and largely constant background in
the spectral distribution.
Consequently, they do not impact the profile of the CUR, which is of
paramount interest.
In the current simulations, this background has not been accounted for.

The results of the spectra calculations have demonstrated that the
intensity and the profile of CUR is sensitive to the value of
the coefficient $\kappa$, which determines the dependence of the
lattice constant on the boron concentration,
Eq. (\ref{Methodology:eq.05}).
The data available in the literature do not permit a definitive choice
of its value.
In order to address this issue, the atomistic-level molecular dynamics
simulations are currently being conducted on boron-doped diamond
crystal.
The objective of this research is to elucidate the effects of dopant
concentration on the lattice constant.
Previously, such simulations were performed in Ref.
\cite{DickersEtAl:EPJD_v78_77_2024}
for silicon-germanium superlattices.
%

\appendix

\section{Periodic part of the profiles due to the
"cosine" and "saw" doping
\label{Profiles}}

\begin{figure} [h]
\centering
\includegraphics[clip,width=11cm]{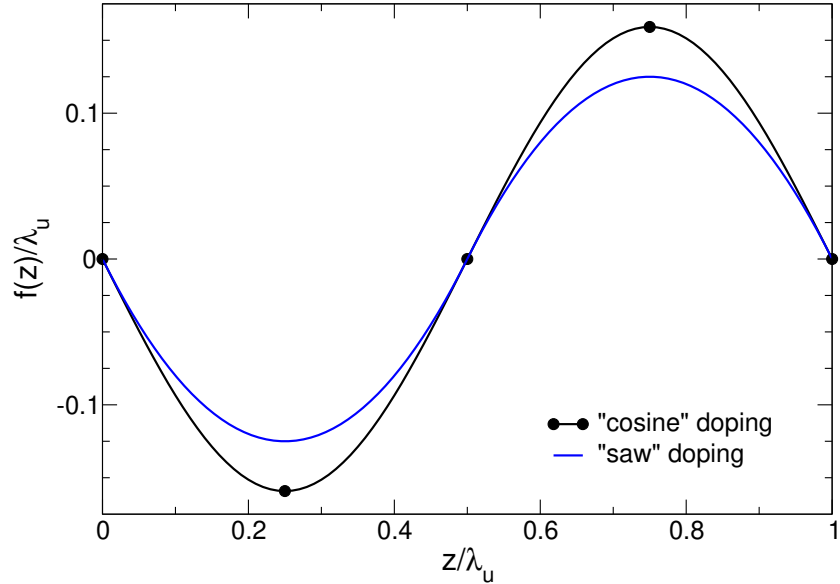}
\caption{
Periodic parts of the profile (\ref{CaseStudies_new:eq.11})
calculated for the "cosine", Eq. (\ref{CaseStudies:eq.01a}),
and the "saw", Eq. (\ref{CaseStudies:eq.01c}),  doping.
}
\label{Figure09.fig}
\end{figure}

Figure \ref{Figure09.fig} compares the functions $f(z)$,
see Eq. (\ref{CaseStudies_new:eq.11}), calculated
for the "cosine" (solid line with filled circles) and "saw"
(solid line) doping schemes.
In the figure, both the abscissa and the ordinate are scaled by the
undulator period $\lamu$.
In these units, the amplitude values are $1/2\pi$ and $1/8$ for
the "cosine" and "saw" doping, respectively.

\section*{Acknowledgements}

We acknowledge support by the European Commission
through the N-LIGHT Project within
the H2020-MSCA-RISE-2019 call (GA 872196)
and the EIC Pathfinder Project TECHNO-CLS
(Project No. 101046458).
We also acknowledge the Frankfurt Center for Scientific
Computing (CSC) for providing computer facilities
We are grateful to Hartmut Backe, Werner Lauth, and  Thu Nhi Tran Caliste
for constructive discussions.
We thank Werner Lauth for careful reading the manuscript.

\vspace*{0.3cm}
\noindent
\textbf{Competing Interests.}
The authors do not declare any conflicts of interest, and there is no
financial interest to report.

\vspace*{0.3cm}
\noindent
\textbf{Author Contribution.}
\textbf{AVK:}
Conceptualization;
Rel-MD simulations;
Analysis of the boron-doped diamond geometry and the results of
atomistic simulations.
Writing – review \& editing.
\\
\textbf{AVS:} Project administration;
Conceptualization;
Analysis of the boron-doped diamond geometry and the results of
atomistic simulations.
Writing – review \& editing.

All authors reviewed the final manuscript.


\section*{References}


\end{document}